\documentclass[a4paper,fleqn]{cas-sc}

\usepackage[numbers,compress]{natbib}

\usepackage{graphicx}
\usepackage{bm}
\usepackage[utf8]{inputenc}
\usepackage[T1]{fontenc}
\usepackage{amsmath}

\usepackage[]{hyperref}
\hypersetup{
 colorlinks = true,
  citecolor = blue,
  urlcolor = red
}
\newcommand{\R}{\mathcal{R}}

\newcommand{\bb}{\bm{b}}
\newcommand{\bj}{\bm{j_\parallel}}

\newcommand{\bu}{\bm{u}_\perp}

\newcommand{\bnabla}{\bm{\nabla}}
\newcommand{\bnablap}{\bm{\nabla}_\perp}
\newcommand{\br}{\bm{r}}

\newcommand{\kperp}{k_\perp}
\newcommand{\wz}{\bm{\omega}}
\newcommand{\jz}{\bm{j}}
\newcommand{\cross}{\bm{\times}}
\newcommand{\la}{\left\langle}
\newcommand{\ra}{\right\rangle}
\newcommand{\lbracket}{\left[}
\newcommand{\rbracket}{\right]}

\def\tsc#1{\csdef{#1}{\textsc{\lowercase{#1}}\xspace}}
\tsc{WGM}
\tsc{QE}

\begin{document}
\let\WriteBookmarks\relax
\def\floatpagepagefraction{1}
\def\textpagefraction{.001}

\shorttitle{}    

\shortauthors{}  

\title [mode = title]{Determination of turbulent heating rate and relaxed states in finite Larmor radius magnetohydrodynamic turbulence with helicity barrier}  



%

\author[]{Ramesh Sasmal}[orcid=0009-0005-5156-0133]



\ead{rameshs21@iitk.ac.in}


\credit{Formal analysis, Writing- original draft, Writing- review and editing, Interpretation}

\affiliation[]{organization={Indian Institute of Technology Kanpur},
            city={Kanpur},
            postcode={208016}, 
            state={Uttar Pradesh},
            country={India}}

\author[]{Supratik Banerjee}[orcid=0000-0002-3746-0989]


\ead{sbanerjee@iitk.ac.in}


\credit{Conceptualization, Validation, Writing- review and editing, Writing- original draft, Supervision}


\cormark[1]
\cortext[1]{Corresponding author}



\begin{abstract}
Finite Larmor radius magnetohydrodynamics (FLR-MHD) provides a hybrid model of plasma that explains how turbulent energy cascade extends to sufficiently small parallel length scales, potentially leading to perpendicular heating of the ions in the solar corona and the solar wind. In this work, we derive exact laws for the cascades of energy and generalized helicity in fully developed FLR-MHD turbulence. In large and small scale limits, we obtain the exact laws for reduced MHD and electron reduced MHD turbulence respectively. Unlike ordinary or reduced MHD turbulence, a global stationary state is shown to be absent in the case of a strong imbalance between the Els\"asser variables. This is due to the so-called helicity barrier, which leads to two separate stationary energy cascades with different cascade rates. Our derived exact laws enable us to calculate these two cascade rates and therefore their difference, which effectively provides the heating rate of the ions. In addition, we also derive alternative Banerjee-Galtier forms for the exact laws and hence obtain the relaxed states of FLR-MHD turbulence using the framework of recently proposed principle of vanishing nonlinear transfer. The relaxed states show alignment between the velocity and magnetic field fluctuations. However, due to strong anisotropy, no Beltrami alignment is possible for velocity and magnetic fields. Similarly to the exact laws, the relaxed states of reduced and electron reduced MHD emerge in the large and small scale limits, respectively.  
\end{abstract}




\maketitle

\section{Introduction}\label{}
Turbulence is a highly nonlinear flow regime characterized by multiscale fluctuations of the flow variables (\textit{e.g.} density, fluid velocity, pressure, \textit{etc.}). Despite its enormous complexity, turbulence is omnipresent in nature and is crucial for understanding efficient mixing, structure formation, and heating of various natural fluid environments. In plasmas, turbulence is characterized in terms of the fluctuations of both plasma variables and electromagnetic fields. For length-scales much larger than the characteristic ion scales i.e. ion inertial scale ($d_i$), ion Larmor radius ($\rho_i$) etc. , a plasma can be modeled as a single magnetohydrodynamic (MHD) fluid. Similarly to ordinary fluids, fully developed turbulence in an MHD fluid also shows inertial scale universal cascades for the inviscid invariants (energy, cross-helicity and magnetic helicity) where the corresponding cascade rate is expressed in terms of velocity and magnetic field fluctuations. However, the presence of a background magnetic field ($\bm{B_0}$) distinguishes between these two types of turbulence. Unlike hydrodynamics, where the Galilean transformation can eliminate the effect of the mean velocity, the effect of mean magnetic field cannot be eliminated from the MHD equations. In the strong anisotropic limit, the turbulent flow features are often described in the framework of reduced magnetohydrodynamics (RMHD) \citep{kadomtsev_1973, strauss_1976}, where the fluctuations permeate and subsequently fragment within the plane perpendicular to the mean field while being advected along it with a speed proportional to ${B_0}$. Consequently, the fluctuation length-scale along the mean field ($l_\parallel$) becomes considerably larger than the fluctuation scale perpendicular to it ($l_\perp$), or equivalently $k_\parallel\ll k_\perp$ in terms of the parallel and perpendicular wavenumbers. 

Despite being developed to study the turbulence in tokamak plasmas, RMHD has also been employed to elucidate several features of space and astrophysical plasmas, such as solar wind turbulence, coronal heating, reconnection, \textit{etc.} \citep{milano_1999, oughton_2001, rappazzo_2007, dalena_2014}. Of these, the anomalous heating of the solar corona and the subsequent acceleration of the solar wind remain the most intriguing problem of space plasma physics. Observational evidence indicates preferential heating of heavier ions in the direction perpendicular to the mean magnetic field \citep{hansteen_1995, coleman_1968, cranmer_1999}. This cannot be explained in the framework of RMHD, which is valid at length and time scales much larger than the ion gyroradius and the ion gyro-period ($\ell \gg \rho_i, \omega\ll \Omega_{i} $), respectively, and leads to the strong heating of the electrons in a low $\beta$ plasma through Alfv\'enic turbulence. Inclusion of compressible MHD waves cannot solve the problem either, as it leads to stronger heating parallel to the mean magnetic field \cite{schekochihin_2009}.
A straightforward extension of RMHD to the kinetic regime invokes Landau damping which efficiently heats the electrons parallel to $\bm{B_0}$ \citep{dobrowolny_1985, leamon_1999}. In addition to a partially understood mechanism of stochastic heating, only ion cyclotron resonance (ICR) can adequately explain the selective and perpendicular heating of the ions \citep{marsch_1982, cranmer_1999, kasper_2013, bale_2019}. However, ICR requires an efficient transfer of energy to very small parallel scales ($\sim d_i$) to satisfy the corresponding resonance condition \cite{kennel_1966, isenberg_2011}. This cannot be achieved by typical energy cascades in MHD (both incompressible and compressible) and RMHD flows, where the energy is primarily transferred to smaller perpendicular scales \cite{Horbury2012,  banerjee_2013, banerjee_2016}. 

Only recently has a self-consistent solution to this enigmatic heating problem been proposed using the model of finite Larmor radius magnetohydrodynamic (FLR-MHD) turbulence. It successfully explains the transfer of energy to sufficiently small parallel scales, thereby enabling a selective heating of ions via ICR \cite{meyrand_2021, squire_2022}. In this hybrid framework, the electron population is considered as a fluid, whereas ions are treated kinetically. FLR-MHD is fundamentally derived from gyrokinetics for a low beta plasma ($\beta_i\sim \beta_e\ll1$) under the assumptions of $k_\parallel\ll k_\perp, \ell\sim \rho_i, \omega\ll \Omega_i$. Within this framework, plasma turbulence can be studied across $\rho_i$ scales in the presence of a strong $\bm{B_0}$. In the inviscid limit, FLR-MHD permits two quadratic invariants: the total energy and the generalized helicity. The second invariant becomes prominent in case of strong imbalance between the generalized Els\"asser variables. For $k_\perp \rho_i \ll 1$, we recover the RMHD limit where the generalized helicity becomes the cross helicity which undergoes a direct cascade in fully developed turbulent state. For $ k_\perp \rho_i \gg 1$, on the other hand, FLR-MHD reduces to electron reduced magnetohydrodynamics (ERMHD), where the generalized helicity becomes proportional to the magnetic helicity which exhibits an inverse cascade. At around $k_\perp \rho_i \sim 1$, these counter-cascading helicities create a helicity barrier, where the majority of forward cascading energy gets suppressed, allowing only a very small fraction of energy to continue cascading to subsequent smaller scales \cite{meyrand_2021}. As energy piles up at the barrier, its location is found to gradually migrate to larger scales \cite{meyrand_2021, squire_2022, adkins_2024}, where it causes the amplitude of the dominating Els\"asser variable to increase. To satisfy the condition of critical balance, $k_\parallel$ also increases as $k_\parallel V_A \sim k_\perp Z^+$ where $Z^+$ denotes the dominant Elsässer field and the Alfv\'en velocity $V_A=B_0/\sqrt{\mu_0\rho_0}$ with $\mu_0$ and $\rho_0$ being the free space permittivity and the density, respectively. Finally, when $k_\parallel$ becomes comparable to $d_i^{-1}$, the frequency of the fluctuations $k_\parallel V_A$ approaches $\Omega_i$, allowing heating through ICR. A turbulent FLR-MHD flow indeed shows the generation of small parallel length scales, as indicated by the increase of parallel dissipation relative to perpendicular dissipation \cite{meyrand_2021}. Moreover, the presence of helicity barrier is recently observed in the solar wind through the turbulence transition of the magnetic spectrum \citep{mcintyre_2025}. 

Despite a few numerical and observational studies, a systematic analytical investigation of FLR-MHD turbulence is still awaited. In particular, no exact law for the cascades of the quadratic invariants has been derived. Such exact relations are crucial to accurately determine the ion heating rate due to turbulence in the solar corona and other astrophysical scenarios where the plasma $\beta$ parameter is moderately low. In this paper, we derive two exact laws, one in classical divergence form \cite{politano_1998}, and the other in alternative form \cite{banerjee_2017}, for the cascades of energy and generalized helicity in FLR-MHD turbulence. From the divergence form, we investigate the possibility of a stationary cascade in the inertial range. We also recover the exact laws of RMHD \cite{ramesh_2025a} as a large-scale limit of the derived exact laws, whereas in the small-scale limit, we obtain the exact laws of ERMHD. Using the alternative exact laws along with the universal theory of fluid and plasma relaxation \cite{banerjee_2023, pan_2024}, finally we predict the relaxed states of FLR-MHD turbulence.

This paper is organized as follows: in Sec. (\ref{flr_mhd}), we introduce the governing equations of FLR-MHD, followed by conservation of the total energy and generalized helicity in the inviscid limit. Sec. (\ref{exactlaw}) consists of the detailed derivation of the exact relations in divergence form. This is followed by Sec. (\ref{stationary}), where we show the absence of a stationary state in strongly imbalanced FLR-MHD along with different limits of the obtained exact laws. In Sec. (\ref{alternative}), we obtain the alternative exact law followed by a detailed derivation of relaxed states of FLR-MHD. Finally, in Sec. (\ref{discussion}), we summarize our result and conclude.

\section{FLR-MHD model} \label{flr_mhd}
\subsection{Governing equations}
In FLR-MHD, we assume a mean magnetic field ($\bm{B_0} = B_0 \hat{z}$) which is stronger than the velocity and magnetic field fluctuations and consequently, the parallel fluctuation length scales are larger than the perpendicular ones ($k_\parallel\ll k_\perp$). Both the ion and electron plasma beta parameters are very small ($\beta_i\sim \beta_e\ll 1$) and fluctuations vary sufficiently slowly compared to the ion-gyro frequency ($\omega\ll \Omega_{i,e} $),  thus yielding 
the governing equations of FLR-MHD  as   
\begin{align}
     \left(\frac{\partial}{\partial t}+ \bu\cdot\bnablap\right)\frac{\delta n_e}{n_{0e}}&=-\frac{1}{\mu_0 e n_{0e}}\left(\frac{\partial}{\partial z}+\bb_\perp \cdot \bnablap\right)\bnablap^2 A_\parallel,\label{governing1} \\
      \left(\frac{\partial}{\partial t}+ \bu\cdot\bnablap\right)A_\parallel&=-\frac{\partial \phi}{\partial z}+\frac{ T_{0e}}{e}\left(\frac{\partial}{\partial z}+\bb_\perp \cdot \bnablap\right)\frac{\delta n_e}{n_{0e}},\label{governing2}
\end{align}

\noindent where $\phi$, $A_\parallel$ are electrostatic and vector potentials, respectively,  $\bu = B_0^{-1} \hat{z} \times \bnablap \phi$ represents the velocity fluctuations, the dimensionless field  $\bb_\perp =-B_0^{-1}\hat{z}\times \bnablap A_\parallel$ represents the direction of the fluctuating magnetic field, $\mu_0$ is the free space permeability, $n_{0i,\, 0e}$ are the ion and electron densities, respectively, $Z$ is the atomic number of the ion, and the temperature ratio between ion and electron is given by $\tau\sim T_{0i}/T_{0e}$. By quasi-neutrality, the perturbed electron density ($\delta n_e$) can be related to the electrostatic potential ($\phi$) as
\begin{equation}
    \frac{\delta n_e}{n_{0e}}=\frac{\delta n_i}{n_{0i}}=M(\bnablap^2) \phi,\label{density1}
\end{equation}

\noindent where $M(\bnablap^2) = -(Ze/\tau T_{0e})(1-\Gamma_0)$ with $\Gamma_0= I_0(\alpha)e^{-\alpha}$, $\alpha= -\rho_i^2\bnabla_\perp^2/2$, and $I_0$ being the modified Bessel function. Denoting $\hat{b} = \bb_\perp + \hat{z}$, the governing Eq. (\ref{governing2}) can be written as 
\begin{equation}
    \frac{\partial A_\parallel}{\partial t} = \lbracket-\hat{b}\cdot\bnabla\left(\phi - \frac{T_{0e}}{e}\frac{\delta n_e}{n_{0e}}\right)\rbracket. \label{governing3}
\end{equation}

\subsection{Inviscid invariants}
In FLR-MHD,  the total energy and the generalized helicity are given by $\int E d\tau$ and $\int H d\tau$, respectively, where 
\begin{align}
    E&= \frac{1}{2}\left[\frac{\Omega_i^2\rho_i^2}{2}\left(-\frac{Ze\phi}{\tau T_{0e}}\frac{\delta n_e}{n_{0e}}+\frac{Z}{\tau}\left(\frac{\delta n_e}{n_{0e}}\right)^2\right)+\frac{(\bnablap A_\parallel)^2}{\mu_0 m_i n_{0i}}\right], \\
    H&= \frac{\Omega_i}{\sqrt{\mu_0 m_i n_{0i}}} \frac{\delta n_e}{n_{0e}}A_\parallel.
\end{align}

\noindent Note that, the above expressions are different from those provided in \citep{meyrand_2021} and can be obtained by linking  the generalized Els{\"a}sser potentials ($\Theta_{\bm{k}}^\pm$) with $\hat{\delta n_e}/n_{0e}$ and $\hat{A}$ (for the detailed derivation, see Appendix (\ref{Appendix})). In this section, we show that the total energy and the generalized helicity are two inviscid invariants of FLR-MHD. Using Gauss divergence theorem, and considering $\delta n_e/n_{0e}= n$, one can show 
\begin{align}
    \int \partial_t (\phi\, n ) \, d\tau&= \frac{2}{\mu_0 e n_{0e}} \int \lbracket \bnablap^2 A_\parallel\, \hat{b}\cdot\bnabla \phi\rbracket\, d\tau, \\
    \int \partial_t(n^2)\, d\tau&= \frac{2}{\mu_0 e n_{0e}}\int \lbracket\bnablap^2A_\parallel\,\hat{b}\cdot\bnabla n\rbracket d\tau,\\
    \int \partial_t (\bnablap A_\parallel)^2\, d\tau &= 2\int \lbracket\bnablap^2A_\parallel\, \hat{b}\cdot\bnabla\phi-\frac{T_{0e}}{e}\bnablap^2A_\parallel\,\hat{b}\cdot\bnabla n\rbracket d\tau.
\end{align}
\noindent Combining the above expressions, one can finally show that $\frac{d}{d t}\int E \,d\tau = 0$. To show the conservation of generalized helicity, one obtains  
\begin{align}
    \int A_\parallel\,\partial_t n \, d\tau&= \int \lbracket n\,\bb_\perp\cdot\bnablap\phi\rbracket\, d\tau,\\
    \int n\,\partial_t A_\parallel \, d\tau&= -\int \lbracket n\, \bb_\perp\cdot\bnablap\phi \rbracket d\,\tau,
\end{align}
\noindent where intermediate total divergence terms vanishes due to Gauss divergence theorem. Adding these two terms, finally we obtain $\frac{d}{dt} \int H d\tau=0$.

\section{Derivations of the exact relations}\label{exactlaw}

\subsection{Exact law for total energy}\label{flr-mhde}

In this section, we derive the exact relation corresponding to the energy transfer in homogeneous FLR-MHD turbulence. For that we need to obtain $\partial_t\R_E$, where 
\begin{equation}
     \R_E=\frac{1}{2}\lbracket-\frac{\alpha e}{2T_{0e}}\la\phi' n + n' \phi\ra + \alpha\la nn'\ra+\beta\la\bm{Q}\cdot \bm{Q'}\ra\rbracket\label{energycorr1}
\end{equation}
\noindent represents the two-point energy correlator, unprimed and primed quantities represent the corresponding fields at positions $\bm{x}$ and $\bm{x}+\bm{r}$, respectively, $\langle(\cdot)\rangle$ represents the ensemble average, $\alpha=\frac{\Omega_i^2\rho_i^2}{2}\frac{Z}{\tau}$, $\beta=\frac{1}{\mu_0 m_i n_{0i}}$, and $\bm{Q}=\bnabla_\perp A_\parallel$. However, the 1st and 2nd term of the above Eq. (\ref{energycorr1}) require $\partial_t\phi$ (and also $\partial_t \phi'$) whereas the governing equations provide $\partial_t n $, where $n=M\phi$. Using statistical homogeneity, one can show 
\begin{equation}
    \la \bnablap^2\phi\, \partial_t \phi'\ra=-\bnabla_{\br_\perp}\cdot\la\bnablap\phi\,\partial_t\phi'\ra = -\la \bnablap\phi\cdot\partial_t \bnablap'\phi'\ra=\bnabla_{\br_\perp}\cdot\la\phi\,\partial_t\bnablap'\phi'\ra=\la \phi \,\partial_t \bnablap'^2\phi' \ra,
\end{equation}
\noindent where we have used the fact that $\bnablap\cdot\la(\cdot)\ra=-\bnabla_{\br_\perp}\cdot\la(\cdot)\ra = \bnabla'_{\br_\perp}\cdot\la(\cdot)\ra$. Since $M$ is a series involving $\bnablap^2$, we can write $\la n\,\partial_t\phi'\ra= \la M\phi\,\partial_t\phi'\ra = \la\phi\,\partial_tM'\phi'\ra=\la \phi\,\partial_t n'\ra$ and similarly, $\la n'\partial_t\phi\ra=\la\phi'\partial_t n\ra$.  By straightforward algebra, one can show    
\begin{align}
     &\partial_t\la \phi'n\ra=\la \phi'\, \partial_t n  +  n \partial_t \phi' \ra= \la \phi'\, \partial_t n +  \phi\, \partial_t n' \ra\nonumber\\
     &= \left\langle \phi'\left[ -(\bu\cdot\bnablap) n + \frac{1}{e n_{0e}}(\hat{b}\cdot \bnabla) j_\parallel \right]+\phi \left[ -(\bu'\cdot\bnablap') n' + \frac{1}{e n_{0e}}(\hat{b}'\cdot \bnabla') j_\parallel' \right]\right>\nonumber\\
     &= \bnabla_{\bm{r}_\perp}\cdot\la n\phi'\bu-n'\phi\bu'\ra+ \frac{1}{e n_{0e}}\bnabla_{\bm{r}}\cdot\la j_\parallel' \phi \hat{b}'- j_\parallel \phi'\hat{b}\ra\label{energycorr3},
\end{align}
\noindent where $\hat{b}= \hat{z}+ \bb_\perp$, $j_\parallel= - (\bnablap^2 A_\parallel)/\mu_0$,
\begin{align}
    &\partial_t\la n n'\ra= \la n'\,\partial_t n + n\,\partial_t n'\ra\nonumber\\
    &= \la n' \lbracket -(\bu\cdot\bnablap) n + \frac{1}{e n_{0e}}(\hat{b}\cdot \bnabla) j_\parallel \rbracket + n \lbracket  -(\bu'\cdot\bnablap') n' + \frac{1}{e n_{0e}}(\hat{b}'\cdot \bnabla') j_\parallel' \rbracket\ra\nonumber\\
    &= \bnabla_{\bm{r}_\perp}\cdot\la n n'\bu-n'n\bu'\ra+ \frac{1}{e n_{0e}}\bnabla_{\bm{r}}\cdot\la j_\parallel' n \hat{b}'- j_\parallel n'\hat{b}\ra\label{energycorr5},
\end{align}
\noindent and 
\begin{equation}
    \partial_t\la \bm{Q}\cdot\bm{Q'}\ra = \la \bm{Q} \cdot \partial_t \bm{Q'} + \bm{Q'}\cdot \partial_t\bm{Q}\ra= \la\bnablap A_\parallel \cdot \partial_t (\bnablap'A_\parallel')+ \bnablap' A_\parallel' \cdot \partial_t (\bnablap A_\parallel) \ra.\label{energycorr6}
\end{equation}
\noindent Again, one can write  $\la\bnablap A_\parallel \cdot \partial_t (\bnablap'A_\parallel')\ra= -\la (\bnablap^2 A_\parallel) \,\partial_t A_\parallel'\ra$ and $\la\bnablap' A_\parallel' \cdot \partial_t (\bnablap A_\parallel) \ra= - \la (\bnablap^{2'}A_\parallel')\, \partial_t A_\parallel\ra$. Using these expressions in the Eq.(\ref{energycorr6}), we have 
\begin{align}
    &\partial_t\la \bm{Q}\cdot\bm{Q'}\ra = -\la (\bnablap^2 A_\parallel) \,\partial_t A_\parallel'+(\bnablap^{2'}A_\parallel')\, \partial_t A_\parallel\ra= \mu_0\la j_\parallel \,\partial_t A_\parallel'+ j_\parallel' \, \partial_t A_\parallel\ra\nonumber\\
    &= \mu_0\la j_\parallel\lbracket -\,\hat{b}'\cdot\nabla'\left(\phi' - \frac{T_{0e}}{e}n'\right) \rbracket+ j_\parallel'\lbracket-\,\hat{b}\cdot\nabla\left(\phi - \frac{T_{0e}}{e}n\right)\rbracket\ra\nonumber\\
    &=\mu_0\, \bnabla_{\bm{r}}\cdot\la \left(\phi - \frac{T_{0e}}{e}n\right)j_\parallel' \hat{b}-\left(\phi' - \frac{T_{0e}}{e}n'\right)j_\parallel\hat{b}'\ra\label{energycorr7}.
\end{align}

\noindent Substituting the expressions (\ref{energycorr3}), (\ref{energycorr5}), and (\ref{energycorr7})
in Eq. (\ref{energycorr1}), one obtains the time evolution of the energy correlator as
\begin{align}
    &\partial_t \R_E\\
    &=\frac{1}{2}\left[-\frac{\alpha e}{T_{0e}}\bnabla_{\bm{r}_\perp}\cdot\la n\phi'\bu-n'\phi\bu'\ra   +\alpha \bnabla_{\bm{r}_\perp}\cdot\la n n'\bu-n'n\bu'\ra   \right. \nonumber\\
    &+\left. \bnabla_{\bm{r}}\cdot\la  \frac{\mu_0 \beta T_{0e}}{e}(n'j_\parallel\hat{b}'-nj_\parallel'\hat{b})+\frac{\alpha}{e n_{0e}}( n j_\parallel'\hat{b}'- n'j_\parallel\hat{b})-\frac{\alpha}{T_{0e} n_{0e}}(j_\parallel'\phi\hat{b}'- j_\parallel\phi'\hat{b})+\mu_0 \beta( j_\parallel'\phi\hat{b}- j_\parallel\phi'\hat{b}') \ra \right]\nonumber\\
    &=\frac{1}{2}\left[-\frac{\alpha e}{T_{0e}}\nabla_{\bm{r}_\perp}\cdot\la \delta n\, \delta \phi\, \delta \bu  \ra + \frac{\alpha}{2}\nabla_{\bm{r}_\perp}\cdot\la(\delta n)^2 \,\delta \bu \ra - \frac{\alpha}{e n_{0e}}\nabla_{\bm{r}}\cdot\la \delta n \,\delta j_\parallel \,\delta \hat{b} \ra+\frac{\alpha}{T_{0e}n_{0e}}\nabla_{\bm{r}}\cdot\la \delta j_\parallel\, \delta \phi \,\delta \hat{b}\ra\right]\nonumber\\
    &= \frac{1}{2} \nabla_{\bm{r}_\perp}\cdot \la-\frac{\alpha e}{T_{0e}}\delta n\, \delta \phi \,\delta \bu+\frac{\alpha}{2}(\delta n)^2\,\delta \bu -\frac{\alpha}{e n_{0e}}\delta n \,\delta j_\parallel \,\delta \bb_\perp+\frac{\alpha}{T_{0e}n_{0e}}\delta j_\parallel\, \delta \phi \,\delta \bb_\perp \ra\nonumber\\
    & = \frac{1}{2} \nabla_{\bm{r}_\perp}\cdot \la \alpha\left(\frac{\delta n}{2}-\frac{e \delta \phi}{T_{0e}}\right)\delta n\,\delta\bu-\frac{\alpha }{e n_{0e}}\left(\delta n - \frac{e \delta \phi}{T_{0e}}\right)\delta j_\parallel\,\delta\bb_\perp\ra,
\end{align}
\noindent where we use  $\mu_0 \beta=\alpha/(T_{0e}n_{0e})$, $\delta \hat{b}= \hat{b}'- \hat{b}= \hat{z}+\bb_\perp'-\hat{z}-\bb_\perp= \delta \bb_\perp$ and consequently, $\bnabla_{\br}\cdot\la(\cdot)\delta \hat{b}\ra=\bnabla_{\br_\perp}\cdot\la(\cdot)\delta\bb_\perp\ra$. The terms $\bnabla_{\br_\perp}\cdot\la n\phi'\bu'\ra$ and $\bnabla_{\br_\perp}\cdot\la n'\phi\bu\ra$ vanish under the assumption of homogeneity since $\bnablap \phi\cdot\bu=0$. Note that the above expression is derived by omitting the forcing and dissipation terms. Including those, the above expression can be written as  
\begin{equation}
    \partial_t \R_E
    =\frac{1}{2} \nabla_{\bm{r}_\perp}\cdot \la \alpha \left(\frac{\delta n}{2}-\frac{e \delta \phi}{T_{0e}}\right)\delta n\,\delta\bu-\frac{\alpha}{e n_{0e}}\left(\delta n - \frac{e \delta \phi}{T_{0e}}\right)\,\delta j_\parallel\,\delta\bb_\perp\ra+ \mathcal{F}_E +\mathcal{D}_E \label{energy},
\end{equation}
\noindent where $\mathcal{F}_E$ and $\mathcal{D}_E$ represent the average contributions due to forcing and dissipation,  respectively. Eq. (\ref{energy}) is the first main result of our paper. It provides an exact law for the total energy cascade in FLR-MHD turbulence. Note that obtaining a stationary form corresponding to the exact law given in Eq. (\ref{energy}) is questionable, as we discuss later.

\subsection{Exact law for generalised helicity}\label{flr-mhdh}

\noindent In order to obtain an exact law for the generalized helicity, we need to calculate $\partial_t \R_H$, where
\begin{equation}
    \R_H= \frac{\gamma}{2}\la n'A_\parallel+ n A_\parallel'\ra\label{helicitycorr1}
\end{equation}
\noindent represents the two-point helicity correlator, and $\gamma=\Omega_i/\sqrt{\mu_0 m_i n_{0i}}$. Using the governing equations, one obtains 
\begin{align}
    &\partial_t\la n'A_\parallel\ra= \la n' \partial_t A_\parallel+ A_\parallel \partial_t n'\ra\nonumber\\
    &= \left\langle n'\left[-\hat{b}\cdot\bnabla\left(\phi- \frac{T_{0e}}{e}n\right) \right]+A_\parallel \left[ -(\bu'\cdot\bnablap') n' + \frac{1}{e n_{0e}}\left(\hat{b}'\cdot \bnabla'\right) j_\parallel' \right]\right\rangle\nonumber\\
    &=\bnabla_{\bm{r}}\cdot\la\left(\phi-\frac{T_{0e}}{e}n\right)n'\hat{b}\ra-\bnabla_{\bm{r}_\perp}\cdot\la n'A_\parallel\bu'\ra+\frac{1}{en_{0e}}\bnabla_{\bm{r}}\cdot\la j_\parallel'A_\parallel\hat{b}'\ra,\label{helicitycorr3}
\end{align} 
\noindent  and 
\begin{equation}
    \partial_t\la A_\parallel' n\ra=\la n\partial_tA_\parallel'+ A_\parallel'\partial_t n\ra
    =-\bnabla_{\bm{r}}\cdot\la \left(\phi'-\frac{T_{0e}}{e}n'\right)n\hat{b}'\ra+\bnabla_{\bm{r}_\perp}\cdot\la nA_\parallel'\bu\ra-\frac{1}{en_{0e}}\bnabla_{\bm{r}}\cdot\la j_\parallel A_\parallel'\hat{b}\ra\label{helicitycorr4}.
\end{equation}
\noindent Adding these above expressions, we obtain 
\begin{align}
    &\partial_t\R_H\nonumber\\
    &=\frac{\gamma}{2}\left[\bnabla_{\bm{r}_\perp}\cdot\la nA_\parallel'\bu- n'A_\parallel\bu'\ra+\bnabla_{\bm{r}}\cdot\la(\phi n' \hat{b}- \phi' n \hat{b}')+\frac{T_0e}{e}(n'n\hat{b}' - n n'\hat{b})+\frac{1}{en_{0e}}(j_\parallel' A_\parallel \hat{b}'- j_\parallel A_\parallel'\hat{b}) \ra\right]\label{helicitycorr5}
\end{align}
\noindent Since $\hat{b}=\hat{z}+\bb_\perp$, one can rewrite the 2nd term of the above expression as  
\begin{equation}
    \bnabla_{\bm{r}}\cdot\la \phi n' \hat{b}- \phi' n \hat{b}'\ra
    =\bnabla_{\bm{r}_\perp}\cdot\la\phi n' \bb_\perp- \phi' n \bb_\perp'\ra+ \la\phi \frac{\partial n'}{\partial z'}- n\frac{\partial \phi'}{\partial z'}\ra=\bnabla_{\bm{r}_\perp}\cdot\la n' A_\parallel \bu- nA_\parallel'\bu'\ra,\label{helicitycorr7}
\end{equation}
\noindent where we use $\la\phi \partial_{z'} n'\ra= \la n\partial_{z'}\phi'\ra$, 
 $\bnabla_{\bm{r}_\perp}\cdot\la\phi n'\bb_\perp\ra= \la n'\bnablap\phi\cdot (B_0^{-1}\hat{z}\times \bnabla_\perp A_\parallel)\ra= \bnabla_{\bm{r}_\perp}\cdot\la n' A_\parallel \bu\ra$ and similarly, $\bnabla_{\bm{r}_\perp}\cdot\la\phi' n \bb_\perp'\ra=\bnabla_{\bm{r}_\perp}\cdot\la nA_\parallel'\bu'\ra$. 
\noindent Finally, The last term of Eq. (\ref{helicitycorr5}) is rewritten as 
\begin{equation}
    \bnabla_{\bm{r}}\cdot\la j_\parallel' A_\parallel \hat{b}'- j_\parallel A_\parallel'\hat{b}\ra= \bnabla_{\bm{r}_\perp}\cdot\la j_\parallel' A_\parallel \bb_\perp'- j_\parallel A_\parallel'\bb_\perp\ra,\label{helicitycorr8}
\end{equation}

\noindent as $\la A_\parallel\partial_{z'}j_\parallel' \ra=\la A_\parallel\partial_{z'}(-\bnablap'^2A_\parallel'/\mu_0) \ra=\la j_\parallel\partial_{z'}A_\parallel'\ra$. Substituting the relations (\ref{helicitycorr7}) and (\ref{helicitycorr8}) in Eq. (\ref{helicitycorr5}), we get
\begin{align}
    &\partial_t\R_H\nonumber\\
    &= \frac{\gamma}{2}\left[\bnabla_{\bm{r}_\perp}\cdot\la (nA_\parallel'\bu- n'A_\parallel\bu'+ n' A_\parallel \bu- nA_\parallel'\bu')+ \frac{1}{en_{0e}}(j_\parallel' A_\parallel \bb_\perp'- j_\parallel A_\parallel'\bb_\perp)\ra + \frac{ T_{0e}}{e}\bnabla_{\bm{r}}\cdot\la n'n\hat{b}' - n n'\hat{b}\ra\right]\nonumber\\
    &= \frac{\gamma}{2}\left[ \bnabla_{\bm{r}_\perp}\cdot\la \delta n\,\delta A_\parallel\, \delta \bu-\frac{T_{0e}}{e}\frac{(\delta n)^2}{2}\delta \bb_\perp- \frac{1}{en_{0e}}\delta j_\parallel\,\delta A_\parallel\, \delta \bb_\perp\ra\right],\label{helicity27}
\end{align}

\noindent where the terms $\bnabla_{\br_\perp}\cdot\la j_\parallel A_\parallel'\bb_\perp'\ra$ and $\bnabla_{\br_\perp}\cdot\la j_\parallel' A_\parallel\bb_\perp\ra$ vanish under the assumption of homogeneity since $\bb_\perp\cdot\bnablap A_\parallel=0$. Finally, in the presence of helicity injection and dissipation, one can write
\begin{equation}
    \partial_t\R_H=\frac{\gamma}{2}\left[ \bnabla_{\bm{r}_\perp}\cdot\la \delta n\,\delta A_\parallel\, \delta \bu-\frac{T_{0e}}{e}\frac{(\delta n)^2}{2}\delta \bb_\perp- \frac{1}{en_{0e}}\delta j_\parallel\,\delta A_\parallel \,\delta \bb_\perp\ra\right]+ \mathcal{F}_H+ \mathcal{D}_H,\label{helicity}
\end{equation}

\noindent where $\mathcal{F}_H$ and $\mathcal{D}_H$ represent the average contributions of helicity injection and dissipation, respectively. Eq. (\ref{helicity}) is the second main result of our paper and it provides an exact law for the generalized helicity cascade in FLR-MHD turbulence.

\section{Large and small scale limits}\label{limit}

\subsection{ At large-scales ($k_\perp\rho_i\ll1$)}
In this section, we obtain different limits of the exact laws derived in the last section. 
At large-scales, we have $k_\perp\rho_i\ll1$ and therefore, $1-\Gamma_0\simeq -\rho_i^2\bnablap^2/2$. In this limit, the Eq. (\ref{energy})  reduces to
\begin{align}
    &\partial_t \R_E\huge {|}_{large} \nonumber\\
    &= \frac{1}{2} \bnabla_{\bm{r}_\perp}\cdot\la - \frac{1}{B_0^2}\delta(\bnablap^2\phi)  \,\delta\phi \,\delta\bu + \frac{1}{m_i n_{0i}}\delta j_\parallel\, \delta\phi\, \delta \bb_\perp\ra+ \mathcal{F}_E +\mathcal{D}_E\nonumber \\
    &= \frac{1}{2}\bnabla_{\bm{r}_\perp}\cdot\la\frac{1}{B_0^2}\left(\bnablap'^2\phi' \phi \bu'-\bnablap^2\phi \phi' \bu\right)+\frac{1}{m_i n_{0i}}\left(-j_\parallel \phi'\bb_\perp'+j_\parallel'\phi\bb_\perp-j_\parallel'\phi\bb_\perp'+j_\parallel\phi'\bb_\perp\right)\ra+ \mathcal{F}_E +\mathcal{D}_E .\label{rmhde3}
\end{align}

\noindent Now, using homogeneity, one can show 
\begin{align}
    &\bnabla_{\bm{r}_\perp}\cdot\la\bnablap'^2\phi'\phi\bu'\ra= -\la(\bnablap'^2\phi')\bnablap\phi\cdot\bu'\ra
    =-B_0^2\la\bu\cdot(\bnablap'\cross\bu')\cross\bu'\ra=-B_0^2\bnabla_{\bm{r}_\perp}\cdot\la(\bu\cdot\bu')\bu'\ra,\nonumber\\
    &\bnabla_{\bm{r}_\perp}\cdot\la-\bnablap^2\phi\phi'\bu\ra=B_0^2\bnabla_{\bm{r}_\perp}\cdot\la(\bu\cdot\bu')\bu\ra\nonumber,\\
    &\bnabla_{\bm{r}_\perp}\cdot\la-j_\parallel\phi'\bb_\perp'\ra=-(B_0^2/\mu_0)\bnabla_{\bm{r}_\perp}\cdot\la(\bb_\perp\cdot\bb_\perp')\bu'-(\bb_\perp\cdot\bu')\bb_\perp'\ra\nonumber,\\
    &\bnabla_{\bm{r}_\perp}\cdot\la j_\parallel'\phi\bb_\perp\ra=(B_0^2/\mu_0)\bnabla_{\bm{r}_\perp}\cdot\la(\bb_\perp\cdot\bb_\perp')\bu-(\bb_\perp'\cdot\bu)\bb_\perp\ra\nonumber,\\
    &\bnabla_{\bm{r}_\perp}\cdot\la j_\parallel\phi' \bb_\perp\ra=-(B_0^2/\mu_0)\bnabla_{\bm{r}_\perp}\cdot\la(\bb_\perp\cdot\bu')\bb_\perp\ra\nonumber,\,\,\text{and}\\
    &\bnabla_{\bm{r}_\perp}\cdot\la -j_\parallel'\phi \bb_\perp'\ra=(B_0^2/\mu_0)\bnabla_{\bm{r}_\perp}\cdot\la(\bb_\perp'\cdot\bu)\bb_\perp'\ra.\nonumber
\end{align}

\noindent Using these relations, the exact law (\ref{rmhde3}) is finally expressed as 
\begin{equation}
     \partial_t \R_E\huge {|}_{large}=\frac{1}{4}\bnabla_{\br_\perp}\cdot\la \left((\delta \bu)^2+\frac{\left(\delta \bm{B}_\perp\right)^2}{\mu_0 m_i n_{0i}}\right)\delta \bu -2 \left(\delta \bu\cdot\delta\bm{B}_\perp\right)\frac{\delta\bm{B}_\perp}{\mu_0 m_i n_{0i}}\ra+ \mathcal{F}_E +\mathcal{D}_E, \label{rmhde4}
\end{equation}

\noindent where $\bm{B}_\perp = B_0\,\bb_\perp$. This relation (\ref{rmhde4}) is similar to the exact law obtained for total energy cascade in RMHD turbulence \cite{ramesh_2025a}. Similarly, at large-scales, the Eq. (\ref{helicity}) can be written as
\begin{align}
   & \partial_t \R_H\huge{|}_{large}\nonumber\\
   &= \frac{\gamma}{2} \bnabla_{\bm{r}_\perp}\cdot\la \delta \left(\frac{Ze \rho^2_i}{2\tau T_{0e}}\bnablap^2\phi\right)\delta A_\parallel\, \delta \bu- \frac{1}{en_{0e}}\delta j_\parallel\,\delta A_\parallel\, \delta \bb_\perp\ra+ \mathcal{F}_H+ \mathcal{D}_H\nonumber\\
   &=\frac{1}{2\sqrt{\mu_0 m_i n_{0i}}}\bnabla_{\bm{r}_\perp}\cdot\la \frac{1}{B_0}\delta (\bnablap^2\phi)\,\delta A_\parallel\, \delta \bu- \frac{B_0}{m_i n_{0i}}\delta j_\parallel\,\delta A_\parallel\, \delta \bb_\perp\ra+ \mathcal{F}_H+ \mathcal{D}_H\nonumber\\
   &=-\frac{1}{4}\bnabla_{\br_\perp}\cdot\la \left(
   (\delta \bu)^2+\frac{\left(\delta \bm{B}_\perp\right)^2}{\mu_0 m_i n_{0i}}\right)\frac{\delta \bm{B}_\perp}{\sqrt{\mu_0 m_i n_{0i}}} -2 \left(\delta \bu\cdot\frac{\delta\bm{B}_\perp}{\sqrt{\mu_0 m_i n_{0i}}}\right)\delta \bu \ra+ \mathcal{F}_H+ \mathcal{D}_H\label{rmhdh3}.
\end{align}

\noindent The above relation is similar to the exact law obtained for the cross helicity cascade in RMHD turbulence \cite{ramesh_2025a}.

\subsection{ At small-scales ($k_\perp \rho_i \gg1$)}
In this limit, we have $1-\Gamma_0 \simeq 1$ and hence, the Eq. (\ref{energy}) reduces to
\begin{align}
    &\partial_t\R_E\huge{|}_{small}\nonumber\\
    &= \frac{1}{2} \bnabla_{\bm{r}_\perp}\cdot \left\langle\frac{\alpha e^2}{T_{0e}^2}\frac{Z}{\tau}(\delta \phi)^2 \,\delta \bu+\frac{\alpha e^2 }{2T_{0e}^2} \frac{Z^2}{\tau^2}(\delta\phi)^2\,\delta \bu  +\frac{\alpha}{ n_{0e}T_{0e}}\frac{Z}{\tau}\delta \phi\, \delta j_\parallel\, \delta \bb_\perp+\frac{\alpha}{T_{0e}n_{0e}}\delta j_\parallel\, \delta \phi \,\delta \bb_\perp\right \rangle+ \mathcal{F}_E +\mathcal{D}_E \nonumber\\
    &= \frac{1}{2} \bnabla_{\bm{r}_\perp}\cdot\la \frac{\alpha}{T_{0e} n_{0e}}\left(1+\frac{Z}{\tau}\right) \delta \phi\, \delta j_\parallel \, \delta \bb_\perp\ra+ \mathcal{F}_E +\mathcal{D}_E \nonumber\\
    &= \frac{1}{4}\left(1+\frac{Z}{\tau}\right)\bnabla_{\bm{r}_\perp}\cdot\la\frac{\left(\delta \bm{B}_\perp\right)^2}{\mu_0 m_i n_{0i}}\delta \bu -2 \left(\delta \bu\cdot\delta\bm{B}_\perp\right)\frac{\delta\bm{B}_\perp}{\mu_0 m_i n_{0i}} \ra+ \mathcal{F}_E +\mathcal{D}_E \nonumber\\
    &=\frac{1}{4}\bnabla_{\bm{r}_\perp}\cdot\la -\frac{1}{n_{0e}e}\lbracket\frac{\left(\delta \bm{B}_\perp\right)^2}{\mu_0 m_i n_{0i}}\delta \bm{j}_\perp-2 \left(\delta \bm{j}_\perp\cdot\delta\bm{B}_\perp\right)\frac{\delta\bm{B}_\perp}{\mu_0 m_i n_{0i}} \rbracket\ra+ \mathcal{F}_E +\mathcal{D}_E \label{ermhde}
\end{align}
\noindent where we use similar mathematical steps as for Eq. (\ref{rmhde3}) along with $\bnabla_{r_\perp}\cdot \la (\delta \phi)^2 \delta \bu\ra=0$ and the fact $B_\parallel/B_0 = (1+Z/\tau)\beta_i Z e\phi/(2T_{0i})\Rightarrow \bm{j}_\perp=-(1+Z/\tau)n_{0e}e\,\bu$ with $\beta_i = 2\mu_0n_{0i}T_{0i}/B_0^2$. At  small scales, FLR-MHD reduces to ERMHD \cite{schekochihin_2009, zocco_2011} and 
one can therefore expect the Eq. (\ref{ermhde}) to correspond to the exact law for the energy cascade in ERMHD turbulence. 
At small-scales, $H$ becomes proportional to the magnetic helicity ($\propto-B_\parallel A_\parallel$, where $B_\parallel$ is the parallel magnetic field fluctuation) and we obtain
\begin{align}
    &\partial_t \R_H\huge{|}_{small}\nonumber\\
    &= \frac{\gamma}{2} \bnabla_{\bm{r}_\perp}\cdot\left\langle \delta \left(-\frac{Z}{\tau}\frac{e }{T_{0e}}\phi\right)\delta A_\parallel\, \delta \bu-\frac{T_{0e}}{2e}\left(\delta \left(\frac{Z}{\tau}\frac{e }{T_{0e}}\phi\right)\right)^2\delta \bb_\perp- \frac{1}{en_{0e}}\delta j_\parallel\,\delta A_\parallel\, \delta \bb_\perp\right\rangle+ \mathcal{F}_H+ \mathcal{D}_H\nonumber\\
    &= -\frac{\gamma}{2}\bnabla_{\bm{r}_\perp}\cdot\la\left(1+\frac{Z}{\tau}\right)\frac{Z e}{2\tau T_{0e}}(\delta \phi)^2\, \delta\bb_\perp+\frac{1}{e n_{0e}}\delta j_\parallel\,\delta A_\parallel\, \delta \bb_\perp\ra+ \mathcal{F}_H+ \mathcal{D}_H\nonumber\\
    &= -\frac{\gamma}{2}\frac{Ze B_0^2}{2\tau T_{0e}}\bnabla_{\bm{r}_\perp}\cdot\la\left(1+\frac{Z}{\tau}\right)\frac{(\delta \phi)^2}{B_0^2} \delta\bb_\perp+\frac{\rho_i^2}{m_in_{0i}}\delta j_\parallel\,\delta A_\parallel \,\delta \bb_\perp\ra+ \mathcal{F}_H+ \mathcal{D}_H\nonumber\\
    &=-\frac{1}{2\mu_0 m_i n_{0i}\sqrt{\mu_0 m_i n_{0i}}}\bnabla_{\bm{r}_\perp}\cdot\la \frac{d_i^2}{\left(1+\frac{Z}{\tau}\right)\rho_i^2}(\delta \bm{B}_\parallel)^2 \delta \bm{B}_\perp+(\delta \bm{B}_\perp)^2\delta\bm{B}_\perp\ra+ \mathcal{F}_H+ \mathcal{D}_H,
\end{align}
\noindent where we use $\bnabla_{\bm{r}_\perp}\cdot\la\delta\phi\delta A_\parallel\delta\bu\ra=\bnabla_{\bm{r}_\perp}\cdot\la(\delta\phi)^2\delta\bb_\perp/2\ra$. Unlike the case of three-dimensional MHD, where the magnetic helicity transfer cannot be expressed as a divergence, here the flux for magnetic helicity cascade can be cast as a two-dimensional divergence \cite{podesta_2008}. Using $n = -(Ze/\tau T_{0e})(\phi-\Gamma_0 \phi)$, one rewrites the Eq. (\ref{helicity27}) as 
\begin{align}
    &\partial_t \R_H\nonumber\\
    &= \frac{\gamma}{2} \bnabla_{\bm{r}_\perp}\cdot\left\langle \delta \left(-\frac{Z}{\tau}\frac{e }{T_{0e}}\phi\right)\delta A_\parallel \,\delta \bu-\frac{T_{0e}}{2e}\left(\delta \left(\frac{Z}{\tau}\frac{e }{T_{0e}}\phi\right)\right)^2\delta \bb_\perp- \frac{1}{en_{0e}}\delta j_\parallel\,\delta A_\parallel\, \delta \bb_\perp\right\rangle \nonumber\\
    &+\frac{\gamma}{2}\bnabla_{\bm{r}_\perp}\cdot\left\langle
    \frac{Ze}{\tau T_{0e}}\delta(\Gamma_0\phi)\,\delta A_\parallel\,\delta\bu+\frac{Z^2e}{2\tau^2T_{0e}}\delta\left(\phi-\Gamma_0\phi\right)\,\delta(\Gamma_0\phi)\,\delta\bb_\perp\right\rangle+ \mathcal{F}_H+ \mathcal{D}_H\nonumber\\
    &= \bnabla_{\bm{r}_\perp}\cdot\bm{F}_{MH} + \bnabla_{\bm{r}_\perp}\cdot\bm{F}_{\Sigma}+ \mathcal{F}_H+ \mathcal{D}_H.
\end{align}
\noindent The above expression clearly indicates the coexistence of two contributions in the transfer of $H$. $\bm{F}_{MH}$ represents the transfer of the magnetic helicity, whereas  $\bm{F}_{\Sigma}$ represents the flux contribution from the rest of $H$. As one moves to very small scales ($k_\perp \rho_i\gg1$), only the contribution due to $\bm{F}_{MH}$ survives, whereas that due to $\bm{F}_{\Sigma}$ becomes negligibly small. 

\section{Absence of global stationary state}\label{stationary}

In this section, we investigate the possibility of a global stationary cascade in the presence of helicity injection. The generalized helicity can also be written as 
\begin{align}
    H &=\frac{\Omega_i}{\sqrt{\mu_0 m_i n_{0i}}} n\, A_\parallel= v_{th_i}n \, \frac{A_\parallel}{\rho_i\sqrt{\mu_0 m_i n_{0i}}}\nonumber\\
    &= \frac{1}{4}\left[\left(v_{th_{i}}n + \frac{A_\parallel}{\rho_i\sqrt{\mu_0 m_i n_{0i}}}\right)^2-\left(v_{th_{i}}n - \frac{A_\parallel}{\rho_i\sqrt{\mu_0 m_i n_{0i}}}\right)^2\right]\nonumber\\
    &=\frac{1}{4}\left[z_p^2-z_m^2\right]
\end{align}

\noindent where $v_{th_{i}}$ is the thermal velocity of ion. The balanced turbulence ($H=0$) corresponds to $z_p \sim z_m$ whereas in imbalanced turbulence ($H\neq0$), one dominates the other such that either $z_p \gg z_m$ or $z_p \ll z_m$. Taking $z_p \gg z_m$, one obtains $A_\parallel/(\rho_i\sqrt{\mu_0 m_i n_{0i}})\sim v_{th_i} n \Rightarrow A_\parallel \sim v_{th_i} \rho_i \sqrt{\mu_0 m_i n_{0i}}\,n= \Omega_i \rho_i^2 \sqrt{\mu_0 m_i n_{0i}}\,n$. Substituting this imbalanced turbulence condition, assuming a global statistical stationary state and neglecting dissipation, the exact laws (\ref{energy}) and (\ref{helicity}) can be written as
\begin{align}
    -\varepsilon_E = &\frac{\Omega_i^2\rho_i^2}{4}\frac{Z}{\tau}\bnabla_{r_\perp}\cdot\la\left(\frac{\delta n}{2}-\frac{e \delta \phi}{T_{0e}}\right)\delta n\,\delta\bu+2\left(\delta n-\frac{e}{T_{0e}}\delta \phi\right)\,\delta(\rho_i^2\bnablap^2n)\,\delta\left(\hat{z}\cross\bnablap(1-\Gamma_0)\frac{\phi}{B_0}\right)\ra,\label{state1}\\
    -\varepsilon_H = &\frac{\Omega_i^2\rho_i^2}{2}\bnabla_{r_\perp}\cdot\la(\delta n)^2\,\delta\bu + 2\delta n\, \delta(\rho_i^2\bnablap^2n)\,\delta\left(\hat{z}\cross\bnablap(1-\Gamma_0)\frac{\phi}{B_0}\right)\ra\label{state2}.
\end{align}

\begin{figure}
    \centering
    \includegraphics[width=.7\linewidth]{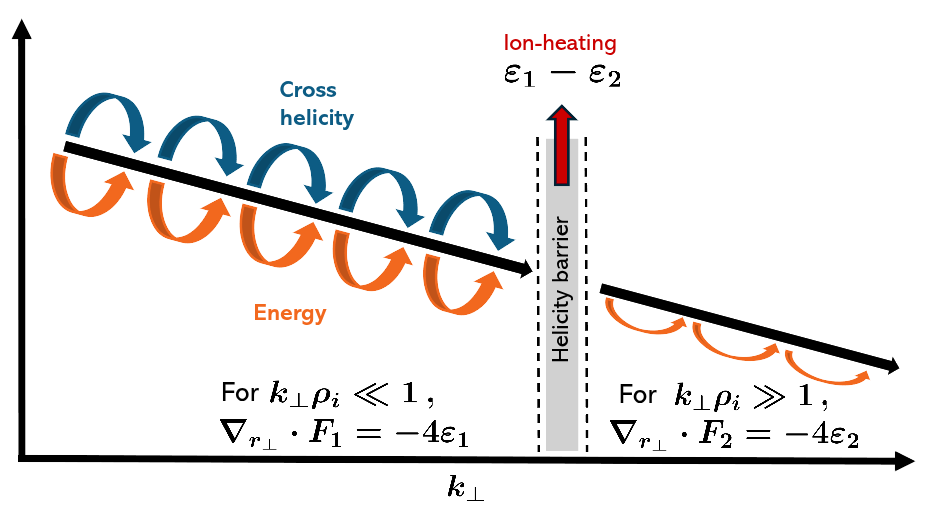}
    \caption{Schematic diagram for ion-heating through helicity barrier mechanism. Here, $F_1$ and $F_2$ denote the flux terms in Eqs. (\ref{eps1}) and (\ref{eps_2}), respectively.}
    \label{fig1}
\end{figure}

\noindent Similar to the Sec. (\ref{limit}), for $k_\perp \rho_i\ll1$, $A_\parallel \sim \Omega_i \rho_i^4 \sqrt{\mu_0 m_i n_{0i}}Ze\bnablap^2\phi/(2\tau T_{0e})$ and the exact laws (\ref{state1})  and (\ref{state2}) take the forms
\begin{align}
    -\varepsilon_E = &\frac{\Omega_i^2\rho_i^2}{2}\bnabla_{r_\perp}\cdot\la-\frac{Ze}{2\tau T_{0e}}\delta\phi \,\delta\left(\frac{Ze}{2\tau T_{0e}}\rho_i^2\bnablap^2\phi\right)\,\delta\bu\ra, \text{and}\\
    -\varepsilon_H = &\frac{\Omega_i^2\rho_i^2}{2}\bnabla_{r_\perp}\cdot\la\delta\left(\frac{Ze}{2\tau T_{0e}}\rho_i^2\bnablap^2\phi\right)\, \delta\left(\frac{Ze}{2\tau T_{0e}}\rho_i^2\bnablap^2\phi\right)\delta\bu\ra,
\end{align}

\noindent whereas for $k_\perp \rho_i\gg1$, $A_\parallel \sim -\Omega_i \rho_i^2 \sqrt{\mu_0 m_i n_{0i}}ze\phi/(\tau T_{0e})$ and the exact laws (\ref{state1}) and (\ref{state2}) become 
\begin{align}
     -\varepsilon_E = &\frac{\Omega_i^2\rho_i^2}{2}\bnabla_{r_\perp}\cdot\la\left(1+\frac{Z}{\tau}\right)\frac{Ze}{\tau T_{0e}}\delta\phi\, \delta\left(\frac{Ze}{\tau T_{0e}}\rho_i^2\bnablap^2\phi\right) \delta\bu\ra, \text{and}\label{helicity_small_E}\\
     -\varepsilon_H = &\frac{\Omega_i^2\rho_i^2}{2}\bnabla_{r_\perp}\cdot\la2\frac{Ze}{\tau T_{0e}}\delta\phi\, \delta\left(\frac{Ze}{\tau T_{0e}}\rho_i^2\bnablap^2\phi\right)\delta\bu\ra.\label{helicity_small_H}
\end{align}
\noindent Since we have assumed global statistical stationarity, we can expect the injection ratio $\sigma=\varepsilon_H/\varepsilon_E$ to be constant in the inertial range irrespective of whether the length scales are larger or smaller than $\rho_i$. However, from the exact laws (\ref{helicity_small_E}) and (\ref{helicity_small_H}), one can observe that $\sigma = 2/(1+Z/\tau)$ which is a constant for small scales, whereas the ratio is not a constant for large scales. A constant ratio at large scale corresponds to $\phi = \rho_i^2\bnablap^2\phi/2$ , thus making the energy and helicity transfer rates to vanish, which is absurd. Therefore, in a statistical stationary state, it is impossible to get a constant $\sigma$ across the entire inertial range. This is a contradiction, and therefore, unlike MHD, in the imbalanced FLR-MHD, we cannot expect stationary cascades for energy and generalized helicity. Numerical studies of FLR-MHD show that counter-cascading cross helicity and magnetic helicity create a barrier (helicity barrier) at length scales of the order of $\rho_i$ \cite{meyrand_2021}. The barrier allows only a very small balanced $2\varepsilon^-$ portion of energy to cascade to scales smaller than $\rho_i$, and the rest of the energy gets accumulated at the barrier with a rate $\varepsilon^+-\varepsilon^-$, where $\varepsilon^\pm=(\varepsilon_E\pm \varepsilon_H)/2$. One can, nevertheless, expect a segment-wise stationary state with different energy and helicity transfer rates if we have a dissipation mechanism at the barrier (parallel dissipation or ion-cyclotron resonance heating) that takes away the energy accumulated at the barrier. If we assume the energy transfer rate to be $\varepsilon_1$ ($\mathcal{F}_E$ is assumed to be $\varepsilon_1$ in stationary state) for $k_\perp \rho_i\ll1$ whereas $\varepsilon_2$ for $k_\perp \rho_i\gg1 $, we can write
\begin{align}
    -4 \varepsilon_1 &= \bnabla_{\br_\perp}\cdot\la \left((\delta \bu)^2+\frac{\left(\delta \bm{B}_\perp\right)^2}{\mu_0 m_i n_{0i}}\right)\delta \bu -2 \left(\delta \bu\cdot\delta\bm{B}_\perp\right)\frac{\delta\bm{B}_\perp}{\mu_0 m_i n_{0i}}\ra \,\,\, \text{for} \,\, k_\perp\rho_i \ll1, \label{eps1}\\
    -4 \varepsilon_2 &= \bnabla_{\bm{r}_\perp}\cdot\la -\frac{1}{n_{0e}e}\lbracket\frac{\left(\delta \bm{B}_\perp\right)^2}{\mu_0 m_i n_{0i}}\delta \bm{j}_\perp-2 \left(\delta \bm{j}_\perp\cdot\delta\bm{B}_\perp\right)\frac{\delta\bm{B}_\perp}{\mu_0 m_i n_{0i}} \rbracket\ra \,\,\, \text{for}\,\, k_\perp\rho_i\gg1.\label{eps_2}
\end{align}
\noindent Now, if we assume the ion-cyclotron resonance heating to be the dissipation mechanism, from the above relations, we can calculate the ion heating rate to be $\varepsilon_1-\varepsilon_2$ (see figure (\ref{fig1})). This is how these exact laws can be employed to accurately determine the ion heating rate in the coronal solar wind.

\section{Alternative exact laws and relaxed states of FLR-MHD turbulence}\label{alternative}

In this section, we derive a Banerjee-Galtier form \citep{banerjee_2017} of the exact relation for FLR MHD turbulence. The evolution equation of the total energy correlation function can be expressed as  
\begin{align}
    &\partial_t \R_E\nonumber\\
    &=\frac{1}{2}\lbracket-\frac{\alpha e}{2T_{0e}}\,\, \partial_t\la\phi' n + n' \phi\ra + \alpha \,\, \partial_t\la nn'\ra + \beta\,\, \partial_t\la\bm{Q}\cdot \bm{Q'}\ra\rbracket\nonumber\\    
    &= \frac{\alpha e}{2T_{0e}} \la \phi'(\bu\cdot\bnablap n)+\phi (\bu'\cdot\bnablap' n')\ra + \frac{\alpha}{2}\la -n'(\bu\cdot\bnablap n) -n(\bu'\cdot\bnablap' n') \ra \nonumber\\
    &+ \la \frac{\alpha}{2T_{0e}n_{0e}}\lbracket -\phi' (\bb_\perp\cdot \bnabla_\perp j_\parallel) -\phi (\bb_\perp'\cdot \bnabla'_\perp j_\parallel')\rbracket + \frac{\mu_0 \beta}{2}\lbracket -j_\parallel'(\bb_\perp\cdot\bnabla_\perp\phi)- j_\parallel(\bb_\perp'\cdot\bnabla'_\perp\phi')\rbracket\ra\nonumber\\
    &+ \la\frac{\alpha}{2 n_{0e} e}\lbracket n'(\bb_\perp\cdot \bnabla_\perp j_\parallel)+ n(\bb'_\perp\cdot \bnabla'_\perp j_\parallel')\rbracket + \frac{\mu_0 \beta T_{0e}}{2e}\lbracket j_\parallel' (\bb_\perp\cdot\bnabla_\perp n) + j_\parallel(\bb_\perp'\cdot\bnabla'_\perp n')\rbracket\ra\nonumber\\
&=\frac{\alpha e B_0}{2T_{0e}}\la\delta\bu\cdot\delta(\bm{\Omega}\times\bu)\ra+\frac{\alpha}{2}\la\delta(\bnabla_\perp\times\bm{\Omega})\cdot\delta(\bm{\Omega}\times\bu)\ra-\frac{\alpha B_0}{2T_{0e}n_{0e}}\la\delta\bu\cdot\delta(\bj\times\bb_\perp)\ra\nonumber\\
    &+ \frac{\mu_0 \beta B_0}{2}\la\delta\bj\cdot\delta(\bb_\perp\times\bu)\ra+\frac{\alpha}{2n_{0e}e}\la\delta(\bnabla_\perp\times\bm{\Omega})\cdot\delta(\bb_\perp\times\bj)\ra- \frac{\mu_0 \beta T_{0e}}{2e}\la\delta\bj\cdot\delta(\lbracket\bnabla_\perp\times\bm{\Omega}\rbracket\times\bb_\perp)\ra,\label{energy_exact 1}
\end{align}

\noindent where we use the following relations 
\begin{align}
    &\la \phi' (\bu\cdot\bnabla_\perp n)\ra = B_0\la (\bu\,\times\, \bm{\Omega})\cdot\bu'\ra, \nonumber \\
    -&\la \phi'(\bb_\perp\cdot\bnabla_\perp j_\parallel)\ra=  B_0\la(\bj\times\bb_\perp)\cdot\bu'\ra\nonumber,\\
   - &\la n'(\bu\cdot\bnabla_\perp n)\ra =  \la(\bu\times\bm{\Omega})\cdot(\bnabla_\perp'\times\bm{\Omega}')\ra\nonumber,\\
    &\la n'(\bb_\perp\cdot\bnabla_\perp j_\parallel)\ra = \la(\bj\times\bb_\perp)\cdot(\bnabla_\perp'\times\bm{\Omega}')\ra\nonumber,\\
    -&\la j_\parallel'(\bb_\perp\cdot\bnabla_\perp \phi)\ra=B_0\la(\bu\times\bb_\perp)\cdot\bj'\ra\nonumber,\\
    &\la j_\parallel'(\bb_\perp\cdot\bnabla_\perp n)\ra= \la\lbracket(\bnabla_\perp\times\bm{\Omega})\times\bb_\perp\rbracket\cdot\bj'\ra\nonumber
\end{align}
with $\bm{\Omega}= n\hat{z}$, and  $\bj= j_\parallel\hat{z}$ and for an arbitrary scalar $m$, we have 
$$\la m' (\hat{z}\cdot \bnabla j_\parallel) +m (\hat{z}\cdot \bnabla' j_\parallel')+j_\parallel'(\hat{z}\cdot\bnabla m) +j_\parallel(\hat{z} \cdot\bnabla' m') \ra=0 .$$

\noindent Similarly, for the generalized helicity, one can have the evolution of the helicity correlation function as 
\begin{align}
     &\partial_t\R_H \nonumber\\
     &=\frac{\gamma}{2}\left\langle-\lbracket n'(\bb_\perp\cdot\bnabla_\perp\phi)+n (\bb_\perp'\cdot\bnabla_\perp'\phi')\rbracket+\frac{T_{0e}}{e}\left[n'(\bb_\perp\cdot \bnabla_\perp n)+n(\bb_\perp'\cdot\bnablap'n')\right]-\lbracket A_\parallel'(\bu\cdot\bnablap n)+A_\parallel(\bu'\cdot\bnablap'n')\rbracket\right.\nonumber\\
     &\hspace{10pt}+\frac{1}{en_{0e}}\left.\left[A_\parallel'(\bb_\perp\cdot\bnablap j_\parallel)+A_\parallel(\bb_\perp'\cdot\bnablap' j_\parallel')\right]\right\rangle\nonumber\\
    &= \frac{\gamma}{2}\la B_0\delta\bm{\Omega}\cdot\delta(\bb_\perp\times\bu)+\frac{T_{0e}}{e}\delta(\bb_\perp\times\bm{\Omega})\cdot\delta(\bnablap\times\bm{\Omega})+B_0\delta\bb_\perp\cdot\delta(\bm{\Omega}\times\bu)-\frac{B_0}{en_{0e}}\delta\bb_\perp\cdot\delta(\bj\times\bb_\perp)\ra \label{helicity_exact 1},
\end{align}
\noindent where we use the following mathematical relations  
\begin{align}
    -&\la n'(\bb_\perp\cdot\bnablap\phi)\ra= B_0\la\bm{\Omega}'\cdot(\bu\times\bb_\perp)\ra\nonumber,\\
    &\la n'(\bb_\perp\cdot\bnablap n)\ra= \la(\bm{\Omega}\times\bb_\perp)\cdot(\bnablap'\times\bm{\Omega}')\ra\nonumber,\\
    -&\la A_\parallel'(\bu\cdot\bnablap n)\ra= B_0\la\bb'_\perp\cdot(\bu\times\bm{\Omega})\ra\nonumber,\\
    &\la A_\parallel'(\bb_\perp\cdot\bnablap j_\parallel)\ra= B_0\la \bb'_\perp\cdot(\bj\times\bb_\perp)\ra\nonumber.
\end{align} along with $ \la n'\hat{z}\cdot\bnabla n+n\hat{z}\cdot\bnabla'n'\ra=0, \, \la n'\hat{z}\cdot\bnabla\phi+n\hat{z}\cdot\bnabla'\phi'\ra=0, \text{ and  }\la A'_\parallel\hat{z}\cdot\bnabla j_\parallel+A_\parallel\hat{z}\cdot\bnabla' j_\parallel'\ra=0$. To prove the last two equations, we use the commutation of $M \text{ and }\bnablap^2$ with $\partial_z$.

\paragraph*{\textbf{Large-scale limit:}} For $k_\perp \rho_i\ll1$, $M \simeq (Ze\rho_i^2/2T_{0i}) \bnablap^2$ and $\bm{\Omega} = (ZeB_0\rho_i^2/2T_{0i})\,\,\wz$ with $\wz= \bnablap\times\bu$. The expressions (\ref{energy_exact 1}) and (\ref{helicity_exact 1}) therefore take the form 
\begin{align}
    \partial_t\R_E\huge{|}_{large} 
    & = \frac{1}{2}\la \delta\bu\cdot\lbracket\delta(\wz\times\bu)+\frac{1}{m_i n_{0i}}\delta(\bm{B}_\perp\times\bj)\rbracket-\frac{1}{m_i n_{0i}}\delta\bj\cdot\delta(\bu\times\bm{B}_\perp)\ra,\label{large1}\\
    \partial_t\R_H\huge{|}_{large}&=\frac{1}{2}\la \frac{\delta\wz}{\sqrt{\mu_0 m_i n_{0i}}}\cdot\delta(\bm{B}_\perp\times\bu)+ \frac{\delta \bm{B}_\perp}{\sqrt{\mu_0 m_i n_{0i}}}\cdot\left[\delta(\wz\times\bu)- \frac{1}{m_i n_{0i}}\delta(\bj\times\bm{B}_\perp)\right]\ra.\label{large2}
\end{align}

\noindent Similarly to the divergence form, these expressions exactly match the alternative exact relations obtained in RMHD turbulence \cite{ramesh_2025a}.

\paragraph*{\textbf{Small-scale limit:}} For $k_\perp\rho_i\gg1$, $M \simeq -Ze/T_{0i} $, $\bm{\Omega}= -\frac{B_0}{n_{0i}T_{0i}\mu_0(1+\frac{Z}{\tau})}\,B_\parallel\,\hat{z}$, and $\bm{j}_\perp=-(1+Z/\tau)n_{0e}e\,\bu$. The expressions (\ref{energy_exact 1}) and (\ref{helicity_exact 1}) are reduced to
\begin{align}
    \partial_t\R_E\huge{|}_{small}&= -\frac{1}{2}\la\frac{1}{m_i n_{0i}}\frac{1}{e n_{0e}}\delta\bm{j}_\perp\cdot\delta(\bm{B}_\perp\times\bj)+\frac{1}{m_i n_{0i}}\frac{1}{e n_{0e}}\delta\bj\cdot\delta(\bm{B}_\perp\times\bm{j}_\perp)\ra,\label{exact_small 1}\\
    \partial_t\R_H\huge{|}_{small} &=-\frac{1}{2}\frac{1}{m_i n_{0i}\sqrt{\mu_0 m_i n_{0i}}}\la\frac{d_i^2}{\left(1+\frac{z}{\tau}\right)\rho_i^2} \delta \bm{B_\parallel}\cdot\delta(\bm{j}_\perp \cross\bm{B}_\perp)+\delta \bm{B}_\perp\cdot\delta(\bj\times\bm{B}_\perp)\ra.\label{exact_small 2} 
\end{align}
\noindent These are the alternative exact relations for the transfer of total energy and magnetic helicity in ERMHD turbulence. Interestingly, both transfers are governed by the nonlinear interaction of current density and magnetic fields.

\paragraph*{\textbf{Finding the relaxed states using PVNLT:}}
We now investigate the states to which a fully developed FLR-MHD turbulence relaxes when the forcing is quenched. For that, we use the recently developed principle of vanishing nonlinear transfer (PVNLT) \cite{banerjee_2023}.  
From equations (\ref{energy_exact 1}) and (\ref{helicity_exact 1}), the average scale-to-scale nonlinear transfer rates for the total energy and generalized helicity are given by 
\begin{align}
    &\la\mathcal{F}^E_{tr}\ra\nonumber\\
    &= -\la\bu'\cdot\frac{\alpha B_0}{2T_{0e}}\left[e(\bm{\Omega}\cross\bu)+\frac{1}{n_{0e}}(\bb_\perp\cross\bj)\right]+\bu\cdot\frac{\alpha B_0}{2T_{0e}}\left[e(\bm{\Omega}'\cross\bu')+\frac{1}{n_{0e}}(\bb_\perp'\cross\bj')\right]\ra\nonumber\\
    &+\la\bj'\cdot\frac{\mu_0\beta}{2}\left[-B_0(\bb_\perp\cross\bu)+\frac{T_{0e}}{e}((\bnablap\cross\bm{\Omega})\cross\bb_\perp)\right]\ra+\la\bj\cdot\frac{\mu_0\beta}{2}\left[-B_0(\bb_\perp'\cross\bu')+\frac{T_{0e}}{e}((\bnablap'\cross\bm{\Omega}')\cross\bb_\perp')\right]\ra\nonumber\\
    &-\la(\bnablap'\cross\bm{\Omega}')\cdot\frac{\alpha}{2}\left[(\bm{\Omega}\cross\bu)+\frac{1}{en_{0e}}(\bb_\perp\cross\bj)\right]+(\bnablap\cross\bm{\Omega})\cdot\frac{\alpha}{2}\left[(\bm{\Omega}'\cross\bu')+\frac{1}{en_{0e}}(\bb_\perp'\cross\bj')\right]\ra\\
    &\la\mathcal{F}^H_{tr}\ra\nonumber\\
    &=-\frac{\gamma}{2}\la B_0\left[\bm{\Omega}'\cdot(\bb_\perp\cross\bu)+\bm{\Omega}\cdot(\bb_\perp'\cross\bu')\right]+\frac{T_{0e}}{e}\left[(\bnablap'\cross\bm{\Omega}')\cdot(\bb_\perp\cross\bm{\Omega})+(\bnablap\cross\bm{\Omega})\cdot(\bb_\perp'\cross\bm{\Omega}')\right]\ra\nonumber\\
    &-\frac{\gamma}{2}\la \bb_\perp'\cdot B_0\left[(\bm{\Omega}\cross\bu)+\frac{1}{en_{0e}}(\bb_\perp\cross\bj)\right]+\bb_\perp\cdot B_0\left[(\bm{\Omega}'\cross\bu')+\frac{1}{en_{0e}}(\bb_\perp'\cross\bj')\right]\ra,
\end{align}
\noindent respectively. According to PVNLT, at relaxed states, both the nonlinear transfers vanish \textit{i.e.} $\la \mathcal{F}_{tr}^{E,\, H}\ra=0$. Trivial relaxed states correspond to $\bu=\bm{0}$ and $\bb_\perp=\bm{0}$. For $\bu= \hat{z}\cross\bnablap\phi=\bm{0}\implies\phi=\phi(z)$ for which $\bm{\Omega}$ vanishes and similarly $\bb_\perp=\hat{z}\cross\bnablap A_\parallel=\bm{0}\implies A_\parallel=A_\parallel(z)$ for which $\bj$ vanishes. Here, we consider general non-trivial relaxed states for which none of $\bu$, $\bb_\perp$, $\bm{\Omega}$, and $\bj$ vanishes. In such cases, the relaxed states require 
\begin{align}
    & (i)\,\,\bb_\perp\cross\bu=\bnablap \psi_1,\label{cond1}\\
    &(ii)\,\, \bb_\perp\cross\bm{\Omega} = \bnablap \psi_2,\label{cond2}\\
    &(iii) \,\, \bm{\Omega}\cross\bu+\frac{1}{en_{0e}}\bb_\perp\cross\jz_\parallel = \bnablap \psi_3, \text{ and }\label{cond3}\\
    &(iv)\,\,B_0(\bu\cross\bb_\perp)+\frac{T_{0e}}{e}(\bnablap\cross\bm{\Omega})\cross\bb_\perp=\bnablap \psi_4\label{cond4}.
\end{align}

\noindent Note that $\bb_\perp\cross\bu$ and $(\bnablap\cross\bm{\Omega})\cross\bb_\perp$ are directed along $\hat{z}$, whereas by definition, $\bnablap \psi_1$ and $\bnablap \psi_4$ must lie in the perpendicular plane. This is only possible when each of $\bb_\perp\cross\bu$ and $(\bnablap\cross\bm{\Omega})\cross\bb_\perp$ vanishes. Therefore, the corresponding relaxed state is characterized by an alignment of $\bu$, $\bb_\perp$, and $\bnablap\cross\bm{\Omega}$. Using this condition in Eqs. (\ref{cond2}) and (\ref{cond3}), we obtain $(\bnablap \cross\bm{\Omega})\cross\bm{\Omega}= \bnablap \psi$ where $\psi$ is another arbitrary scalar function. Note that this would never reduce to a  Beltrami alignment condition for $\bm{\Omega}$ as $\bm{\Omega}$ is directed along $\hat{z}$ whereas $\bnablap\cross\bm{\Omega}$ remains confined in $x-y$ plane. This simply indicates $\bnablap \psi$ cannot vanish, and for the same reason, $\bnablap \psi_2$ and $\bnablap \psi_3$ (in Eqs. (\ref{cond2}) and (\ref{cond3}) respectively) cannot vanish either. The final states of relaxation of FLR-MHD turbulence are therefore given by
\begin{align}
    & (i)\,\,\bb_\perp\cross\bu=\bm{0},\label{cond5}\\
    &(ii)\,\, \bb_\perp\cross\bm{\Omega} = \bnablap \psi_2,\label{cond6}\\
    &(iii) \,\, \bm{\Omega}\cross\bu+\frac{1}{en_{0e}}\bb_\perp\cross\jz_\parallel = \bnablap \psi_3, \text{ and }\label{cond7}\\
    &(iv)\,\,(\bnablap\cross\bm{\Omega})\cross\bb_\perp=\bm{0}\label{cond8}.
\end{align}

\paragraph*{\textbf{Relaxed states at large and small scale limits:}} At large scale, $k_\perp\rho_i\ll1$ and hence $\bm{\Omega} \approx (ZeB_0\rho_i^2/2T_{0i})\,\,\wz$. Using this, one can show $\vert (T_{0e}/e)(\bnablap\cross\bm{\Omega})\cross\bb_\perp\vert/\vert B_0(\bu\cross\bb_\perp)\vert \sim k_\perp^2\rho_i^2\ll1$ and hence, the second term of Eq. (\ref{cond4}) can be neglected with respect to the first term. As a result, Eq. (\ref{cond1}) and (\ref{cond4}) practically become similar. In the large scale limit, Eq. (\ref{cond2}) reduces to $\bb_\perp\cross\wz=\bnablap\tilde{\psi_2}$ whereas Eq. (\ref{cond3}) becomes $\wz\cross\bu+(B_0/m_i n_{0i})\bb_\perp\cross\bj=\bnablap \tilde{\psi_3}$. 
Using similar arguments as above, one can finally obtain the relaxed states as
\begin{equation}
    \bm{B}_\perp\cross\bu=\bm{0} \text{ and } \wz\cross\bu+\frac{1}{m_i n_{0i}}\bm{B}_\perp\cross\bj=\bnablap \tilde{\psi_3},\label{relax_rmhd}
\end{equation}
\noindent where $\bm{B}_\perp=B_0\,\bb_\perp$. As expected, these states exactly match the non-trivial relaxed states obtained in \cite{ramesh_2025a}.

For small scale, $k_\perp\rho_i\gg1$ and hence, $\bm{\Omega}\approx -(Ze/T_{0i})\phi \hat{z}$ and $\bm{j}_\perp=-(1+Z/\tau)n_{0e}e\,\bu$. Therefore, in this limit, Eq. (\ref{cond1}) and (\ref{cond4}) become practically similar. Again, Eq. (\ref{cond2}) reduces to $\bb_\perp \cross\phi \hat{z}=\bnablap \psi_2'$ whereas Eq. (\ref{cond3}) becomes $\bb_\perp\cross\bj= \bnablap \psi_3'$. Using the fact that $\phi \hat{z}$ and $\bj$ are directed along the same direction, $\bb_\perp \cross\phi \hat{z}=\bnablap \psi_2'$ practically becomes $\bb_\perp\cross\bj= \bnablap \Psi$, where $\Psi$ is scalar function. Accumulating all the conditions, one finally obtains the relaxed states as 
\begin{equation}
    \bm{B}_\perp\cross\bu=\bm{0}\text{ or, }\bm{B}_\perp\cross\bm{j}_\perp=\bm{0}, \text{ and } \bm{B}_\perp\cross\bj= \bnablap \Psi_1.\label{relax_ermhd}
\end{equation}
\noindent These correspond to the relaxed states of ERMHD turbulence. Note that, instead of taking different limits of Eqs. (\ref{cond1})-(\ref{cond4}), one can directly obtain the relaxed states given in Eqs. (\ref{relax_rmhd}) and (\ref{relax_ermhd}) from the alternative exact laws Eqs. (\ref{large1})-(\ref{exact_small 2}) obtained for RMHD and ERMHD turbulence.

\section{Discussion}\label{discussion}

FLR MHD is an important hybrid model for space plasmas that can consistently explain the occurrence of ion cyclotron resonance as a consequence of turbulent cascading of energy.  In this paper, we derive exact laws for total energy and generalized helicity cascade in homogeneous three-dimensional FLR-MHD turbulence. Similarly to MHD and RMHD turbulence, it can be tempting to obtain a single exact relation for each cascade by assuming a global stationary cascade with a constant cascade rate \cite{politano_1998, ramesh_2025a}. However, such a globally stationary cascade is shown to be impossible for strongly imbalanced FLR-MHD turbulence due to the presence of helicity barrier, where the helicity cascade stops and the energy cascade only allows a very small portion of energy to cascade to smaller scales. Although argued previously using approximative arguments associated with numerical support \cite{meyrand_2021}, we analytically justify this fact in this present work. Instead, at large times, we can expect a segment-wise (scales larger and smaller than $\rho_i$) stationary cascade if we have a dissipation mechanism near the helicity barrier (supposedly due to parallel dissipation or ion-cyclotron resonance heating). If the accumulated energy at helicity barrier gets dissipated through ion heating, the corresponding heating rate can be given by $\varepsilon_1-\varepsilon_2$ (see figure (\ref{fig1})). Owing to the derived exact laws, this heating rate can therefore be determined in terms of two point fluctuations of plasma and electromagnetic field variables. Numerical simulations are being developed to show the co-existence of this double energy cascade and calculate the associated proton heating rate. In addition, we have also shown different limits of the FLR-MHD exact relations. For $k_\perp\rho_i\ll1$, we have recovered  the exact law for the transfer of total energy and cross helicity in RMHD turbulence \cite{ramesh_2025a}, whereas for $k_\perp\rho_i\gg1$, we get the exact relations for energy and magnetic helicity cascade in ERMHD turbulence. Note that, the exact laws for ERMHD have not been derived till date.     

Using the divergence form, we can also predict energy spectra in stationary FLR-MHD turbulence. Though we cannot have a global stationarity for imbalanced FLR-MHD turbulence, we can still predict energy spectra if there is a segment-wise stationary cascade. For $k_\perp \rho_i\ll1 $, assuming $u_\perp \sim B_\perp/\sqrt{\mu_0 m_i n_{0i}}\Rightarrow \phi \sim B_0 A_\parallel/\sqrt{\mu_0 m_i n_{0i}}$, from the exact relation (\ref{rmhde4}), one can get the energy spectra $\mathcal{E}(k_\perp, k_\parallel)\sim k_\perp^{-5/3}k_\parallel^{-1}$ for axisymmetry satisfying the constitutive relation $3\alpha_\perp+2\alpha_\parallel=7$ \cite{galtier_2005, ramesh_2025b}. Relaxing the axisymmetric condition, from Boldyrev's phenomenology, one can also obtain perpendicular $-3/2$ spectra \cite{Boldyrev_2005, boldyrev_2006, boldyrev_2009}. Similarly, for $k_\perp \rho_i \gg 1$, assuming $\phi\sim (T_{0e}/eB_0)\, k_\perp A_\parallel$ and axisymmetric turbulence, one obtains $\mathcal{E}(k_\perp, k_\parallel)\sim k_\perp^{-7/3}k_\parallel^{-1}$. 

In addition to the usual divergence form, we have also derived the alternative exact relations for energy and generalized helicity. Similarly to the divergence form, for $k_\perp\rho_i\ll1$, we recover the alternative form of exact laws obtained in RMHD turbulence, whereas for $k_\perp\rho_i\gg1$, we have obtained the alternative form of the exact laws in ERMHD turbulence. Using PVNLT, we obtain the states in which a fully developed FLR-MHD turbulence relaxes if we withdraw the forcing. The relaxed states of FLR MHD are shown to exhibit an alignment between $\bu$ and $\bm{B}_\perp$. However, due to the anisotropy of the system, FLR MHD turbulence never relaxes to a state of Beltrami alignment since $(\bnablap\cross\bm{m})\cross\bm{m} \neq \bm{0}$, where $\bm{m}$ is a vector either parallel to the mean magnetic field or confined in a plane perpendicular to it. Therefore, FLR MHD turbulence relaxes to a state with a finite pressure gradient such as $(\bnablap\cross\bm{\Omega})\cross\bm{\Omega}=\bnablap \psi$. For $k_\perp \rho_i \ll1 $, the obtained states reduce to the relaxed states of RMHD, whereas for $k_\perp\rho_i \gg1$, we obtain the relaxed states of ERMHD. The existence of the obtained relaxed states can also be numerically verified and is taken as a future project.  

The derived exact relations can be employed to precisely measure the ion heating rate of the solar wind using \textit{in-situ} high-resolution data with a presence of a strong mean magnetic field and a gradient length-scale sufficiently smaller than $\rho_i$. However, implementation of the \textit{in-situ} data in the derived exact law is non-trivial and can be an important step forward in solar wind turbulence. This work can also be extended to account for finite electron inertia, which is represented in the framework of the kinetic reduced electron heating model (KREHM). Finally, one can also study the effect of intermittency in FLR MHD turbulence and how this small-scale intermittency influences large-scale intermittency, resulting in dynamic alignment \cite{mallet_2016, schekochihin2022}.

\printcredits

\appendix
\section{Appendix}\label{Appendix}
In terms of Els\"asser potential ($\Theta^\pm_{\bm{k}}$), total energy and generalized helicity densities are given by 
\begin{equation}
    \mathcal{E}= \frac{1}{4}\sum_{\bm{k}} \left(|\kperp \Theta^+_{\bm{k}}|^2 + |\kperp \Theta^-_{\bm{k}}|^2 \right)\,\, \text{and}\,\,
    \mathcal{H}=\frac{1}{4}\sum_{\bm{k}} \frac{1}{v_{ph}(\kperp)}\left(|\kperp \Theta^+_{\bm{k}}|^2 - |\kperp \Theta^-_{\bm{k}}|^2\right),\label{app1}
\end{equation}
\noindent where 
\begin{equation}
     \Theta^\pm_{\bm{k}} = -\Omega_i \frac{v_{ph}(k_\perp)}{\kperp^2}\hat{\frac{\delta n_e}{n_{0e}}} \mp \frac{\hat{A_\parallel}}{\mu_0 m_i n_{0i}}\,\, \text{and}\,\, v_{ph}(k_\perp)= \frac{\kperp \rho_i}{\sqrt{2}}\left(\frac{1}{1-\hat{\Gamma_0}}+\frac{Z}{\tau}\right)^\frac{1}{2}
\end{equation}
\noindent Substituting these in Eq.(\ref{app1}), one obtains energy density as
\begin{align}
    \mathcal{E}&=\frac{1}{2}\sum_{\bm{k}}\left(\frac{\Omega_i^2 v_{ph}^2(k_\perp)}{k_\perp^2}|\hat{n}|^2+\frac{ k_\perp^2|\hat{A_\parallel}|^2}{\mu_0 m_i n_{0i}}\right)\nonumber\\
    &=\frac{1}{2}\sum_{\bm{k}}\left(\frac{\Omega_i^2\rho^2_i}{2}\lbracket\frac{1}{1-\hat{\Gamma}_0}+\frac{Z}{\tau}\rbracket|\hat{n}|^2+\frac{ k_\perp^2|\hat{A_\parallel}|^2}{\mu_0 m_i n_{0i}}\right)\nonumber\\
    &=\frac{1}{2}\sum_{\bm{k}}\left(-\frac{\Omega_i^2\rho_i^2Ze}{2\tau T_{0e}}\hat{\phi}\hat{n}^*+\frac{\Omega_i^2\rho^2_i Z}{2\tau} |\hat{n}|^2+\frac{ k_\perp^2|\hat{A_\parallel}|^2}{\mu_0 m_i n_{0i}} \right)\nonumber\\
    &= \frac{1}{2}\int \frac{d\tau}{V}\left(\frac{\Omega_i^2\rho_i^2}{2}\lbracket-\frac{Ze}{\tau T_{0e}}\phi n+\frac{Z}{\tau} n^2\rbracket+\frac{(\bnablap A_\parallel)^2}{\mu_0 m_i n_{0i}}\right),
\end{align}
\noindent where $\hat{n} = \hat{\delta n_e}/n_{0e}$ and $\hat{n}^*$ represents conjugate of $\hat{n}$. Similarly, generalized helicity density can be given by
\begin{align}
    \mathcal{H} &= \frac{1}{2}\frac{\Omega_i}{\sqrt{\mu_0 m_i n_{0i}}}\sum_{\bm{k}}\left(\hat{n}\hat{A_\parallel}^*+\hat{n}^*\hat{A_\parallel}\right)\nonumber\\
    &= \frac{\Omega_i}{\sqrt{\mu_0 m_i n_{0i}}}\int \frac{d\tau}{V} nA_\parallel.
\end{align}


\bibliographystyle{elsarticle-num}




\end{document}